\documentstyle[aps,eqsecnum]{revtex}
\input epsf
\draft
\twocolumn
\title{Novel Ferromagnetic Atom Waveguide with {\em in situ} Loading}
\author{M. Vengalattore, W. Rooijakkers and M.
Prentiss}
\address{Center for Ultracold Atoms, Lyman Laboratory,\\
Physics Department, Harvard University, Cambridge MA 02138}
\date{\today}
\begin{document}
\maketitle
\begin{abstract}
    Magneto-optic and magnetostatic trapping is realized near a surface using
    current carrying coils wrapped around magnetizable cores. A
    cloud of $10^7$ Cesium atoms is
    created with currents less than 50 mA. Ramping up the current while
    maintaining optical dissipation leads to tightly confined atom clouds
    with an aspect ratio of 1:1000.
    We study the 3D
    character of the magnetic potential and characterize
    atom number and density as a function of the applied current.
    The field gradient in the transverse dimension has been varied
    from $<$ 10 G/cm to $>$ 1 kG/cm. By loading and cooling atoms in-situ, we have
    eliminated the problem of coupling from a MOT into a smaller phase space.
\end{abstract}
\pacs{PACS numbers: 32.80.Pj, 03.75.Be, 39.20.+q, 73.63.Nm}
% \twocolumn
%\narrowtext
%\section{Introduction}
    Encouraged by the recent successful demonstration of Bose
    Einstein condensation in atomic vapors, and the macroscopic
    coherence properties of this new state of matter, many groups
    are pursuing the integration of several complex atom-optical devices
    on a single substrate\cite{hansch1,schmied1,cornell1,prentiss1,hinds1}.
    %One example is a robust reciprocal large area
    %interferometer that could be used as an ultrasensitive
    %gyroscope\cite{pritch1,kasevich1}.
    Atomic waveguides are likely to be a building block of such
    atom-optic devices.
    Magnetic waveguides for atoms have been demonstrated,
    confining weak field seeking atoms in a local magnetic field
    minimum. Most guides are based on current carrying wires,
    either suspended in free space\cite{hansch2,schmied2},
    embedded in a tube\cite{hinds2,raithel} or
    lithographically patterned on a
    substrate\cite{schmied1,cornell1,prentiss2}.

    %In this paper, we describe a new approach to magnetic
    %waveguides eliminating two major difficulties encountered in
    %earlier work.  Atom guiding experiments using hollow core optic fibers,
    %donut-mode laser beams and lithographically fabricated surface
    %traps have attempted to transfer a large number of atoms from a large
    %capture volume to a tightly confining structure.
    %The severe mode mismatch results in a poor transfer efficiency.
    %We demonstrate good atom transfer efficiency between a high
    %aspect ratio MOT and a tightly confining magnetic waveguide by
    %using the same quadrupole field to form the MOT and the
    %initial low gradient magnetic waveguide.
    %Thus, all the atoms are loaded {\em in
    %situ}, avoiding the problem of coupling the atom cloud
    %into a tightly confining waveguide. The atoms transferred
    %into this waveguide are subsequently compressed by increasing
    %the magnetic field gradient by around 3 orders of magnitude\footnote{This 2D
    %trapping, transfer and compression is analogous to the 3D
    %trapping, transfer and compression process used to create most
    %Bose condensates.}.
    In this paper, we describe a new approach to create magnetic
    waveguides that transport neutral atoms above the surface of a
    substrate, by exploiting magnetic materials
    poled by time dependent currents. This approach has several major advantages over
    systems that simply use current carrying wires to create the
    waveguides.

        Firstly, the capture volume for collecting atoms from
        the background vapor can be much larger.
        For a typical gradient of 10 G/cm, the capture volume for
        a wire-based guide is about $10^{-4}$ cm$^2$ times the length
        of the guide.
        In contrast, the ferromagnetic structures used in our experiment
        have capture volumes of around 1 cm$^2$ times the length.
        This is due to the different range of the
        magnetic potential determined by the size of the
        structure, 100 $\mu$m for the wire-based traps and 1 cm in
        our case.
    The large capture volume has
    allowed us to trap a large number of atoms in a magneto-optical trap (MOT)
    using the same quadrupole field that will be
    used to form the magnetic waveguide.
    We transfer the atoms
    from the MOT into the waveguide by ramping the magnetic field by around 3 orders of
    magnitude to produce a very high gradient waveguide
    at the same position as the initial MOT\cite{bose} with the transferred atoms
    remaining cold in all three dimensions. Since all the atoms are loaded {\em in
    situ}, the transfer between the low and high gradient
    waveguides can be made very efficient.  Thus we avoid the more
    traditional process of dropping or pushing atoms into a
    tightly confined waveguide which results in inefficient
    transfer and increase in temperature.

    Secondly, the large distance of the center of the waveguide from the
    surface allows us to protect the atoms from heating due to
    interactions between the atoms and the surface\cite{wilkens}.
    Moreover, the high field gradients result in
    a transverse width of the atom cloud which is much smaller
    than the distance from the surface. This
    inhibits losses due to collisions with the surface\cite{hansch1}.
    Also, any inhomogeneities in the magnetic field due to small imperfections
    on the surface of the ferromagnetic foils are smoothed out at the location of
    the waveguide.

        Thirdly, the amplification of the magnetic field by the
        ferromagnetic foils implies a smaller current is required to
        generate a given peak magnetic field. This results in
        greatly reduced ohmic heating.
        In our setup, the ferromagnetic
        cores amplify the field due to the wires typically by a
        factor of 50. Thus, the initial gradient of
        around 7 G/cm for the MOT was realized by running less
        than 50 mA around the ferromagnetic cores. We have
        achieved gradients on the order of 1 kG/cm and trap depths
        of more than 100 G with less than 100 W of dissipation.
        This reduced heating is important since one of the modes
        of failure of previous devices based on current carrying
        wires was the fusing of wires due to local heating when
        running currents of a few amperes\cite{drndic}.

        Finally, these structures permit much larger field
        gradients and trap depths than can be obtained in systems using only
        current carrying wires. The peak field at the
        surface of a ferromagnetic core that is fully poled is
        equal to the saturation field of that material which
        ranges from around 4 kG to 20 kG\cite{ohandley}. Steep
        magnetic traps can be formed by using ferromagnetic
        structures of suitably small size\cite{vuletic1,vuletic2}.
        This results in a larger
        compression of the atoms from the initial MOT, leading to
        higher densities and improved collision rates for
        evaporative cooling. Also, the large gradients imply
        tighter confinement and larger ground state energies and mode
        spacings.
        Moreover,  this would enable us to guide the confined
    atoms in tight loops permitting the fabrication of reciprocal
    atom interferometers with large enclosed areas and small
    length\cite{pritch1,kasevich1}. For instance, our demonstrated
    gradient of 1 kG/cm would allow us to transport atoms moving
    at 1 m/s around a loop of radius 6 mm giving an interferometer
    with an enclosed area larger than 1 cm$^2$.

        For a quadrupole potential $U(r) =
        F |r|$, the transverse mode spacing is  $\left(
        F^2 \hbar^2 / 2 m \right)^{1/3}$ in energy units and the ground state width is
        given by $\left( 2 m F / \hbar^2\right) ^{-1/3}$.
        We have experimentally achieved 1D magnetic traps with
        transverse mode spacings of around 10 kHz (0.5 $\mu$K) and ground
        state sizes less than 50 nm.
        The large aspect ratio (1:1000) of the eventual magnetostatic trap
        realized in our experiment makes this system an attractive candidate
        for investigation of the Tonk's gas regime\cite{olshanii}.
        Similar structures of smaller size, made with ferromagnetic
        materials of larger saturation fields, should permit
        gradients of $10^7$ G/cm. The configuration described in
        this work can also be used to split a waveguide in the
        transverse dimension with the aid of an external bias
        field.

\begin{figure}
\epsfxsize=3.5in \epsfysize=2.5in \centerline{\epsfbox{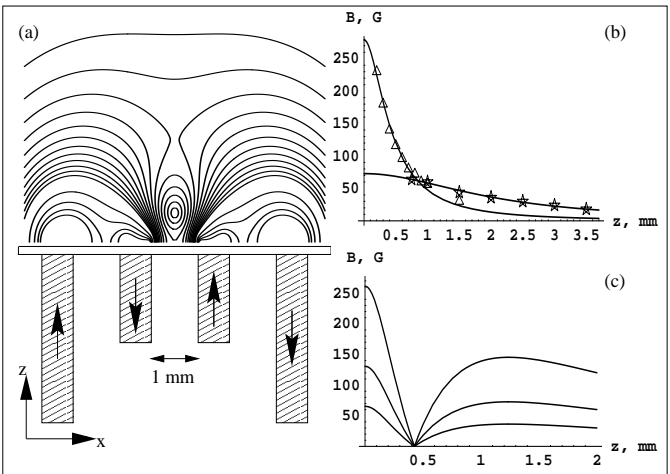}}
\caption{(a) Schematic of the magnetic structure. The arrows on
the ferromagnetic foils indicate the direction of magnetization.
The current around each foil is 1 A. The contours are spaced 20 G
apart. (b) The field of the inner ($\triangle$) and outer pair
($\star$) in the plane of symmetry measured with an audio tape
head. These data fit to a simple model described in the text. (c)
Scaling of the field due to the combined inner and outer pair of
foils in the plane of symmetry. The current around the outer foils
($I_{outer}$) is twice that around the inner foils ($I_{inner}$).
$I_{inner}$ = 0.5, 1 and 2A. } \label{fig1}
\end{figure}

A schematic of the experimental setup is shown in Fig.\ref{fig1}.
Two pairs of ferromagnetic foils (80\% Ni, 15.5\% Fe, 4.5\% Mo)
are mounted on a rigid platform at separations of 1 mm. Both pairs
of foils are 0.5 mm thick and 5 cm long along the long axis. The
inner pair of foils is cut to a height of 1 cm while the outer
pair is cut to a height of 2.5 cm. Thin kapton insulated wires are
wound around these foils with 9 turns around the inner pair and 25
turns around the outer pair. Depending on the sense of the current
around each foil, the foils can be individually magnetized in one
of two directions.
In the situation where the foils are magnetized in opposite
directions, the direction of the field at the plane of symmetry
due to one pair of foils is completely horizontal and its
magnitude varies with height above the surface as\cite{davis}
\begin{equation}
    B(z,x=0) = c_0 \,\, \frac{\mu_0}{2 \pi} \,\, \frac{N I}{l} \,\, \frac{s w}{z^2 + s^2/4 + w^2/4}
\end{equation}
    where $I$ is the current around the ferromagnetic foils, $N$ the number of windings and
    $z$ the height above the surface. $l$, $s$ and $w$ are the
    height of the foils, separation between the pair of foils and the
    thickness of each foil respectively. The dimensionless constant $c_0$
    depends on the geometry and material of the ferromagnetic
    foils and is experimentally found to be around 50.

        Due to the geometry of these foils, the potential is very
        shallow along the long
        (waveguide) axis and the magnetic trap has a large aspect
        ratio.
        A pair of trapping beams perpendicular to the long axis of the
        waveguide is
        reflected off a gold coated mirror at an angle of 45$^\circ$. This mirror is
        glued to the surface of the foils as indicated in
        Fig.\ref{fig1}.
        Another pair of counterpropagating laser beams grazes the
        mirror surface along the long axis of the guide. This
        geometry is similar to a mirror MOT\cite{hansch1} but in
        this case, the quadrupole fields are generated by the
        ferromagnetic foils underneath the mirror. Additionally,
        changing the polarization of the laser beams along the
        guide does not significantly change the number of trapped atoms
        indicating that the atoms are damped but not trapped in
        this dimension. The length of the cloud along the
        waveguide is limited by the waist of the laser beam and
        equals 1 cm in our case.
        We emphasize that the MOT in which the atoms
        are initially loaded and cooled is at the same
        location as the magnetic
        waveguide into which the atoms are eventually loaded
        and compressed.

 It is important to characterize the fields generated by the
ferromagnetic foils as a function of the current and the height
above the surface. This was done by two independent methods.

        In the first method, the fields due to the inner and outer pair of
foils were determined with a calibrated commercial tape head on a
micrometer translation stage. The measurements were performed by
running ac currents of 10 mA at a frequency of 150 Hz around the
foils. It was verified that the pickup voltage does not depend on
the ac frequency. These measurements are shown in Fig.1(b). The
calculations in Fig.1(c) show that the gradient can be increased
while keeping the trap minimum at the same height above the
surface.

    Next, the system was placed in a UHV chamber pumped to a typical
    pressure of $5 \times 10^{-9}$ mbar.
    Currents of up to 1A were run around either the inner or the
    outer pair of foils. By cancelling these fields at the plane
    of symmetry with an external horizontal field of known
    magnitude, magneto-optic traps were created at different
    heights above the surface. At the position of the cloud, the
    magnitude of the bias field is equal to that of the
    ferromagnetic foils, by construction.
    Therefore, this
    constitutes an alternative calibration of the field as a function of
    distance to the surface. This measurement using cold atoms
    shows good agreement with the data in Fig.1(b).
%\begin{figure}
%\epsfxsize=3in \epsfysize=2in
%\centerline{\epsfbox{figures/fig2.eps}} \caption{Measured position
%of the MOT above the surface. The quadrupole field was generated
%using one pair of foils and an external bias field. The position
%of the MOT agrees with the calibration shown in Fig.1(b). This
%also shows that the distance to the surface can be controlled.}
%\label{fig2}
%\end{figure}

    Using an external bias field does not exploit the benefits
    of the ferromagnetic materials and
    imposes a limit on the maximum
    gradient that can obtained.
    However, MOTs can also be formed by exclusively running currents
    around the inner
    and outer pair of foils, without any external bias field. By using the
    outer pair of ferromagnetic foils to provide the cancelling
    field, waveguides with larger gradients and trap depths can be
    realized. Also, the waveguide can be formed without an
    external field.
    When $I_{inner}$ and $I_{outer}$ are increased by the same
    factor $\gamma$, the gradient should increase by $\gamma$
    without changing the height of the trap as shown in Fig.1(c).
    To verify this, we have measured the height of the atom cloud above
    the surface and plotted the results in Fig.2. In this
    experiment, the ratio $I_{outer}/I_{inner}$ were 1/3, 1/2 and
    3/4 corresponding to heights of 1.5 mm, 1.1 mm and 0.8 mm
    respectively. This is in agreement with the previous calibration.

\begin{figure}
\epsfxsize=3in \epsfysize=2in \centerline{\epsfbox{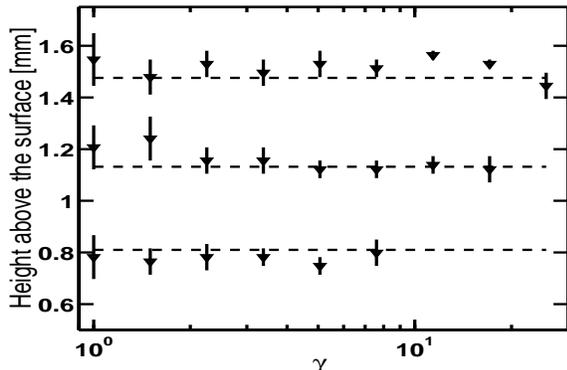}}
\caption{Position of the magneto-optic trap when $I_{inner}$ and
$I_{outer}$ are increased in proportion by a factor $\gamma =
I_{inner}/I_0$ with $I_0 = 150$mA. The dotted lines show the
position of the traps predicted from the calibration using the
atomic clouds (see text). } \label{fig3}
\end{figure}

\begin{figure}[b]
\epsfxsize=3.5in \epsfysize=2.5in
\centerline{\epsfbox{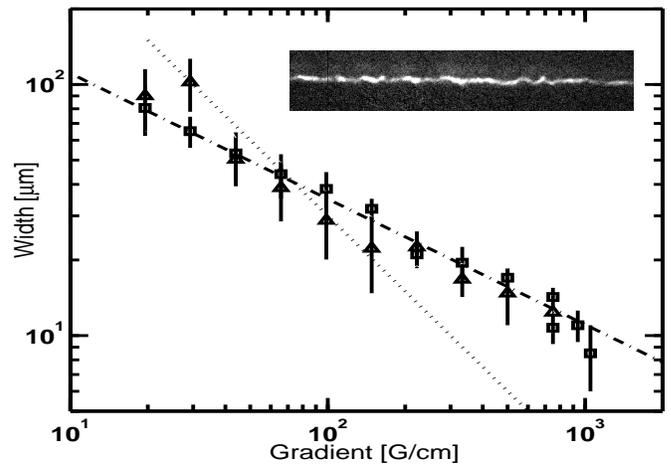}} \caption{Transverse widths(HWHM)
of the MOT($\Box$) and the magnetic trap($\triangle$) vs.
gradient. The dashed and dotted lines indicate slopes of -1/2 and
-1 respectively. Inset: Fluorescence image of a 2.5 mm long
section of the MOT at 330 G/cm.} \label{fig4}
\end{figure}

    Quantitative information about the temperature and the density
    of the cloud requires investigating the width of the MOT.
    Initially, atoms were collected at a gradient of 7 G/cm. Then,
    $I_{inner}$ and $I_{outer}$ were slowly increased in
    proportion. After 20 ms, the cloud was imaged on a CCD camera
    to determine the half width at half maximum (HWHM).
    Assuming that the transverse motion of the atoms is in
    equilibrium with the initial temperature,
     the transverse width of the magneto-optic trap
     $\langle r^2 \rangle \approx k_{B} T/\kappa$
    where $\kappa$, the spring constant of the magneto-optic
    trap is linearly proportional to the gradient $b$. Thus,
    $\langle r \rangle_{MOT} \sim b^{-1/2}$. An
    investigation of the transverse widths of the MOT and the
    magnetic trap for a wide range of gradients (Fig.\ref{fig4})
    shows excellent agreement with this model.
    To our knowledge, this large dynamic range (almost 3 orders of magnitude)
    has not been achieved in a waveguide before.
    We have also investigated the width of
    the cloud in the magnetostatic trap. In this case, the laser beams
    were switched off 3 ms after ramping the currents.
    After a dark period of 100 ms, the MOT beams were switched on
    to image the cloud for 6 ms. For adiabatic compression, we
    expect the width to scale as $b^{-1/3}$ for a linear potential
    and $b^{-1/4}$ for a harmonic potential. However, assuming that the MOT
    beams cool the cloud to its initial temperature, the width
    should scale as $b^{-1}$ for a linear potential.
    As indicated by the dotted line in Fig.3, this is the case in the
    regime 10 $< b <$ 200 G/cm. Above 200 G/cm, the oscillation
    period $T = 2 \pi \sqrt{m/\kappa}$ of atoms in the MOT is
    smaller than the camera exposure time, restoring the MOT width.
    Similar results were obtained when the gradient of the
    magnetostatic trap was increased in the dark.
    Assuming an equal distribution of
    $m_F$ states, the transverse
    temperature is calculated to be approximately 20 $\mu$K.
    These data suggest that
    dissipation can be maintained well into the regime of
    collisional loss\cite{pritch2}.

%This is also an
%indication of the adiabaticity of transfer into the high
%gradient traps.
   % It should be noted that magneto-optic traps at
   % large densities suffer from extremely high rates of light
   % induced collisional losses and heating\cite{pritch2}, which cause a
   % deviation from the behavior described above.

\begin{figure}[b]
\epsfxsize=3.5in \epsfysize=2.5in \centerline{\epsfbox{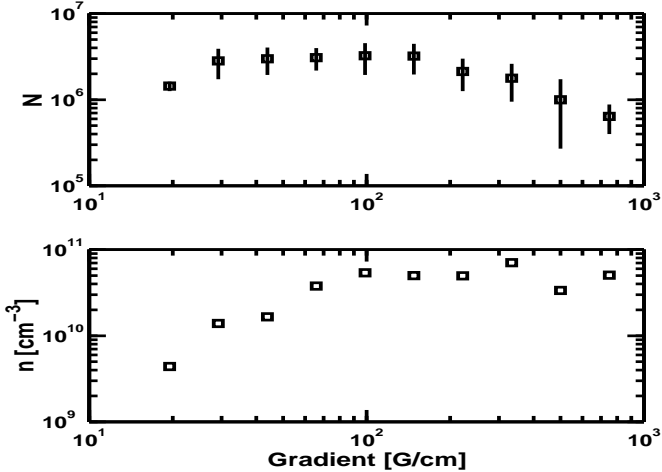}}
\caption{The measured number and peak density of atoms as a
function of the gradient.} \label{fig5}
\end{figure}

The number of atoms in the elongated atom clouds is deduced from
the fluorescence images. Due to the small transverse size of the
atom clouds, a commercial parfocal zoom lens (VZM 450) is used for
magnification. The solid angle of this lens and the efficiency of
the CCD chip were measured in a separate experiment. The
scattering rate of atoms in the MOT was calculated using the
experimental parameters for the laser intensity and detuning,
averaging over the Clebsch Gordon coefficients. The variation of
atom number with the gradient is shown in Fig.\ref{fig5}. This
data can be combined with the transverse width of the atom clouds
to estimate the peak density at the various gradients. It can be
seen that the density saturates at large gradients. This feature
has been observed in 3D traps\cite{cornell2}, and is due to
multiple photon scattering. The number of atoms decreases at the
higher gradients due to light induced collisional losses.
Increasing the detuning and decreasing the intensity of the
trapping light as the gradient is ramped up should improve the
number of trapped atoms.

%
%\section{Conclusion}
    In conclusion, we have demonstrated direct loading of atoms
    into a ferromagnetic waveguide with large gradient and trap depth.
    Our measurements of the magnetic fields, both with an audio tape head and with a cold
    atom cloud, agree well with a simple model. Due to the large
    relative permeability, ferromagnetic materials offer the
    possibility of reaching large energy splittings between
    vibrational states. This large gap may be important to
    preserve coherence during matter wave transportation
    and to facilitate cooling. By
    applying an external bias field in the system described, we
    have also demonstrated a bifurcation of the trapping potential
    to form a double well\cite{wilbert}. This results in an
    experimentally observed
    splitting of the elongated cloud in the transverse dimension
    which is coherent in the adiabatic limit. An atom
    interferometer in time may be formed by applying a
    nonadiabatic phase shift to one of the arms\cite{hinds4}.

        Our system also permits a long guide ($\sim$ 1 m), where
        cold atoms are continuously injected at one end, and
        evaporative cooling is performed along the guide. Dalibard
        and coworkers\cite{dalibard} have proposed a similar system to create
        a cw atom laser.
        Our experiments
        offer prospects of creating Bose condensates on a
        substrate on a much shorter time scale or even continuously.

The authors gratefully acknowledge valuable discussions with N. H.
Dekker, S. A. Lee and  G. Zabow. We are indebted to  T. Deng for
preparing ultrathin gold mirrors. This work was supported by
National Science Foundation Grant Numbers: PHY-9876929 and
PHY-0071311 and MURI Grant Number: Y-00-0007.


\begin{references}
\bibitem{bose}This 2D trapping, transfer and compression is
analogous to the 3D trapping, transfer and compression process
used to create most Bose condensates.
\end{references}
\end{document}